# INTERCONNECT CHALLENGES IN HIGHLY INTEGRATED MEMS/ASIC SUBSYSTEMS

*Norman Marenco, Stephan Warnat, Wolfgang Reinert*

Fraunhofer ISIT, Fraunhoferstrasse 1, D-25524 Itzehoe, Germany

## ABSTRACT

Micromechanical devices like accelerometers or rotation sensors form an increasing segment beneath the devices supplying the consumer market. A hybrid integration approach to build smart sensor clusters for the precise detection of movements in all spatial dimensions requires a large toolbox of interconnect technologies, each with its own constraints regarding the total process integration. Specific challenges described in this paper are post-CMOS feedthroughs, front-to-front die contact arrays, vacuum-compliant lateral interconnect and fine-pitch solder balling to finally form a Chip-Scale System-in-Package (CSSiP).

## 1. INTRODUCTION

Smart products for mobile consumer applications contain an increasing amount of sensor functions. A special field of sensorics consists of inertial measurement units (IMU): Micromechanical elements that generally need to be clustered to allow movement detection in all spatial dimensions. These devices are core elements of exciting and highly promising new product developments for enhanced user interaction and context sensitivity. However, the costs and size of a fully equipped multiaxial sensor unit are still too high for a large scale consumer market introduction.

The DAVID project (Downscaled Assembly of Vertically Integrated Devices) addresses both the cost and size problem by specific research activities on core technologies for hybrid system-in-package (SiP) solutions. The focus lies on waferlevel packaging technologies and chip-to-wafer die bonding.

One of the big challenges in a consequent SiP integration process lies in merging various interconnect technologies: Former approaches basically contained distinct routing levels and directly corresponding technology platforms and -suppliers. New concepts require a high-density threedimensional interconnect structure with severe impacts on design rules and process technology of MEMS and ASIC.

The interconnect elements considered within DAVID are: Post-CMOS through-silicon vias (TSV), a direct front-to-front contact array between ASIC and MEMS, vacuum compliant lateral interconnect on ASIC, and fine-pitch solder balling.

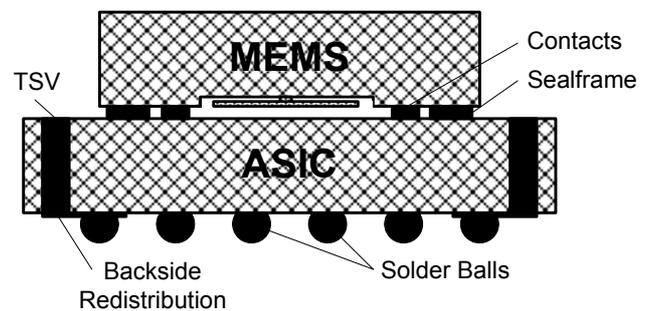

Related technology elements are suitable metallization stacks, passivation technologies and conformal coating in deep etch holes. Effects of process technologies on MEMS functionality and ASIC performance are evaluated by simulation and characterization of a specially designed test vehicle.

## 2. POST-CMOS THROUGH-SILICON VIAS

Inertial sensors are mostly realized as surface micromachined silicon structures that are sensitive to mechanical stress. A stress reduction can be achieved by the present MEMS/ASIC stacking approach. However, an extreme thinning of the silicon wafers like in memory stacks and logic devices is not recommended. In consequence, the TSV has to pass through a relatively thick silicon wafer. To maintain a reasonable contact pitch, a high aspect ratio has to be achieved (e.g. 1:10).

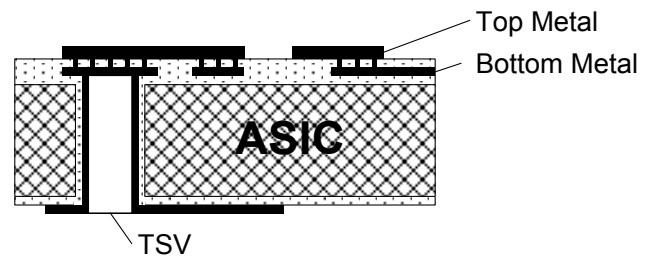

Deep reactive ion etching (DRIE) as a dry etching technology has already produced good results through wafer thicknesses of 300 µm and more. Despite of highly selective etch rates around 1:100, thick etching masks are therefore needed. DRIE causes structural artefacts on the sidewalls that are known as scalloping and notching. They can almost completely be suppressed by a suitable process parameter adjustment, but still remain a risk for the electrical performance and reliability of the TSV.

Various technologies of conformal coating on vertical sidewalls are under investigation for passivation and metallization with CMOS compliant materials and processes. One of the conditions for a successful process integration is that the temperature never exceeds a maximum of 400°C. In case of optical sensors, the limit can even lie below 200°C due to organic coatings on the frontside.

Many processes for oxide and nitride passivation layers do not work under these conditions; however, good results have been achieved with plasma enhanced chemical vapor deposition (PECVD).

Sputtering and evaporation as the most established thin-film metallization processes are generally optimized for planar deposition. For conformal coating in a deep etched hole, process modifications are necessary: Beneath metal-organic CVD, investigations on advanced sputtering technologies are carried out and have already shown very promising results.

An evident impact of TSV implementation on the ASIC design is that no functional structures must be present below the pads. Additionally, the designer has to provide an adequate pad design to avoid voltage offsets within the bottom metal level caused by an inversed vertical supply current flow.

### 3. FRONT-TO-FRONT MEMS/ASIC CONTACT ARRAY

The principal integration approach in DAVID consists of using the ASIC chip as part of a hermetically closed waferlevel package by placing a MEMS directly on top of it. A galvanically deposited sealframe is used to enclose a very thin cavity for the sensitive moving MEMS structures (e.g. a resonating gyroscope or an accelerometer). Electrical contacts are established together with the sealframe and lie within the cavity to allow the closest possible connection between sensor elements and driver circuits. An evident benefit is an improved signal-to-noise ratio of the capacitive sensing. Additionally, signal routing formerly done in polysilicon on MEMS side can be integrated into the ASIC's high-density metal multilayer design.

Electromagnetic crosstalking from the CMOS circuits to the MEMS sensor electrodes can follow from the proximity of the active structures on both dies; adequate shield layers might be needed.

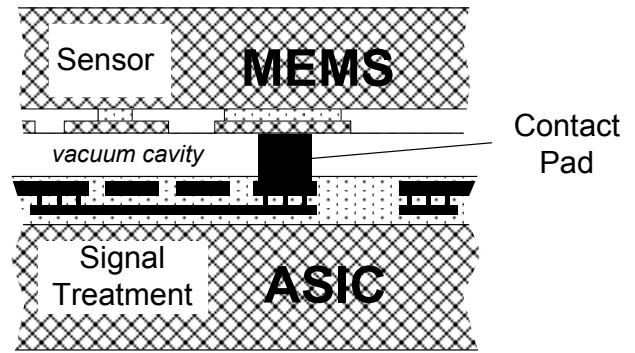

Local mechanical stress from the bonding process acts directly onto the processed structures. Countermeasures against damage and functional deterioration must be developed on design level and in form of technology optimization. A high involvement of CMOS design and integration experts, equipment suppliers and academic partners in the DAVID project shall lead to a set of design rules addressing the potential risks.

### 4. VACUUM COMPLIANT LATERAL INTERCONNECT ON ASIC

One of the integration concepts in DAVID uses a chip-to-wafer (C2W) bonding technology of MEMS on ASIC. It is targeted to use the ASIC's top metal level as the supporting layer for a galvanically realized vacuum sealframe and to provide a contact array inside of it. Consequently, lateral connections from inside to outside must be routed within the underlying metal levels. The vacuum compliance of such a feedthrough is not self-evident due to possible gas permeation, e.g. of helium, through oxide layers.

Mechanical stress on the sealframe area has to be distributed by adequate design measures on ASIC level to avoid cracking of dielectric layers.

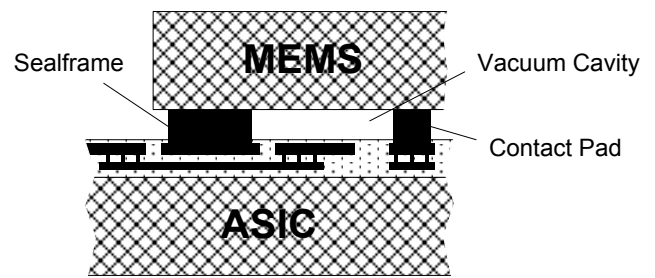

An encapsulation in mold compound shall stabilize the assembly against external mechanical impacts.

Transfer molding is generally made at very high pressures and causes significant local forces on the contact areas. Finite element simulations are carried out in DAVID to detect critical design and process configurations, the result verification is part of test vehicle functions.

## 5. FINE-PITCH SOLDER BALLING

For $2^{nd}$ level interconnect, a solder balling process shall be applied on overmolded wafers. ISIT's standard balling process uses screen printing of flux and a mask-based solder ball placement. Since wafer warpage cannot be completely avoided, difficulties are expected when balls become too small. In DAVID, the limits of ball size and pitch are determined and new concepts to overcome these are investigated.

## 6. SUMMARY AND CONCLUSION

Interconnect technologies for zero-level packaging of stress-sensitive devices have severe impacts on design rules and process technology at all product levels. Mechanical aspects become particularly important due to the fact that a sensor cavity is integrated in the product. The availability of process elements for through-silicon vias is physically limited by the thermal sensitivity of CMOS wafers. Restricted access to frontend equipment for externally processed wafers is one of the infrastructure-bound criteria that has to be considered in each practical realization, in particular in all pre-production phases. Although our technical approach seems principally feasible, high investments in equipment will be necessary for a transfer to the targeted high-volume production scale.

## 7. ACKNOWLEDGEMENTS

The described work is part of the DAVID project, a specific targeted research project funded by the European Commission within the Sixth Framework Programme (contract ref. IST-027240).

Project homepage:     http://www.david-project.eu